\documentclass[sn-mathphys,Numbered]{sn-jnl} 

\usepackage{subcaption}
\usepackage{graphicx}
\usepackage{multirow}
\usepackage{booktabs}
\usepackage{array}
\usepackage{amsmath,amssymb,amsfonts}
\usepackage{amsthm}
\usepackage{mathrsfs}
\usepackage[title]{appendix}
\usepackage{xcolor}
\usepackage{textcomp}
\usepackage{manyfoot}
\usepackage{algorithm}
\usepackage{algpseudocode}
\usepackage{listings}
\usepackage{hhline}
\usepackage{multicol}
\usepackage{enumerate}
\usepackage{natbib}
\usepackage{float}
\PassOptionsToPackage{hyphens}{url} 
\usepackage{hyperref} 

\DeclareUnicodeCharacter{202F}{\,}

\theoremstyle{thmstyleone}

\theoremstyle{thmstyletwo}

\theoremstyle{thmstylethree}

\begin{document}
	
	\title[]{ Testing the weak cosmic censorship conjecture via test particle-induced overcharging/overspinning of Kerr–Newman-modified gravity black hole
	}
	

	\author[1]{\fnm{Waqar} \sur{Ahmad}\textsuperscript{} (waqar.ahmad@cust.edu.pk)}
	\author[2]{\fnm{Abdul Rehman} \sur{Kashif} (kashmology@gmail.com)}
	

	\affil[]{\orgdiv{} \orgname{Capital University of Science and Technology}, \city{Islamabad}, \country{Pakistan}}
	



\abstract{We investigate the weak cosmic censorship conjecture (WCCC) violation through the absorption of a charged, rotating test particle by Kerr-Newman spacetime in modified gravity (KN-MOG). The result is sensitive to multiple factors, including the sign of the particle’s charge relative to that of the black hole, as well as the direction of rotation of the particle and the black hole.
	 Additionally, the combined effect of the MOG parameter, the black hole's charge and angular momentum plays a significant role. Taking all these into account, we have determined the range of the particle's charge, angular momentum and energy of the particle for which the event horizon disappears. Our analysis shows that the WCCC can be violated in both extremal and non-extremal KN-MOG black hole, provided the particle’s parameters are small and precisely adjusted.
	Crucially, prior work by \cite{khodabakhshi2020cosmic} disussed the WCCC in KN sapcetime without considering the modified gravity parameter $\alpha$. In contrast, our study shows that incorporating $\alpha$ significantly enlarges the range of permissible parameters for the test particle, leading to the violation of the WCCC in both extremal and non-extremal KN-MOG black holes.}

\maketitle

\section{Introduction}\label{intro}  
The phenomenon of gravitational collapse often results in the formation of singularities in spacetime. In general relativity, these singularities are expected to be hidden behind event horizons, thus preventing their observation by distant observers. This expectation is formalized in the cosmic censorship conjecture, originally proposed by Roger Penrose in 1969 \cite{penrose1969gravitational}. Specifically, the weak version of this conjecture \cite{penrose2019singularity} asserts that singularities formed from physically reasonable matter should always remain cloaked by a black hole’s event horizon, precluding the formation of visible—or “naked”—singularities. Although a rigorous proof remains elusive, various studies suggest that singularities arising from generic, smooth initial conditions are indeed likely to remain hidden from external observers \cite{wald1999gravitational}.

To test the robustness of this conjecture, one critical question is whether the event horizon of a black hole can be destroyed. In a thought experiment, Wald \cite{wald1974gedanken} investigated whether sending a test particle into a black hole could lead to the destruction of its event horizon. For a general KN-MOG black hole, the WCCC is commonly examined by adding charged or rotating particles, or perturbative fields, to see if they can eliminate the black hole's event horizon.

Building upon Wald’s foundational study, numerous researchers extended this line of inquiry by analyzing the interaction of test particles with black holes in electro-vacuum spacetimes \cite{hubeny1999overcharging,gao2013destroying,siahaan2016destroying,saa2011destroying,siahaan2017destroying,yu2018cosmic,jacobson2009overspinning,de2001turning,ahmed2020weak,ahmed2020cosmic}. Parallel investigations employed test fields to examine similar effects.  \cite{semiz2011dyonic,duztacs2013cosmic,duztacs2020testing,toth2016weak,natario2016test,sorce2017gedanken}. Subsequent research extended these investigations to asymptotically anti-de Sitter spacetimes, focusing on the behavior of probes and fields within such curved spacetime \cite{ying2020thermodynamics,gwak2016cosmic,gwak2016thermodynamics,gwak2018weak,chen2019thermodynamics}.

The appearance of singularities in gravitational systems suggests that general relativity may break down at very small scales, where quantum gravitational effects are expected to dominate. In this context, Modified Gravity (MOG) theory has emerged as a viable alternative, offering consistent explanations for various astronomical, astrophysical, and cosmological observations. Notably, MOG accounts for key features of the universe, including: (i) the acoustic peaks observed in the cosmic microwave background radiation, (ii) the accelerated expansion indicated by type Ia supernovae, and (iii) the galaxy-galaxy correlation patterns seen in the matter power spectrum \cite{moffat2007modified}.

Liang et al.~\cite{liang2019weak} investigated the possibility of destroying the Kerr-MOG black hole by absorbing a test particle, without considering self-force or radiation effects. By examining the dynamics of particle motion and the stability of the black hole's event horizon, they identified the energy bounds—both minimum and maximum—within which a test particle could induce a transition to a naked singularity. Their analysis revealed that, under specific conditions, the weak cosmic censorship conjecture (WCCC) may be violated for both extremal and near-extremal black holes.

In a related work, Koray~\cite{duztacs2020overspinning} studied the validity of the WCCC with black hole thermodynamics in the context of Kerr-MOG BHs exposed to scalar fields. The study demonstrated that scalar fields with frequencies marginally exceeding the superradiant threshold could transform both extremal and near-extremal Kerr-MOG black holes into naked singularities. This outcome remained valid even when backreaction effects were included.

In an earlier investigation, Koray~\cite{duztacs2019kerr} focused on the interaction between neutral scalar test fields and Kerr-Newman black holes. His analysis demonstrated that extremal Kerr-Newman black holes could, in principle, be overspun, leading to a violation of the WCCC. He argued that backreaction effects might serve as a protective mechanism to preserve the event horizon. Nevertheless, when extending the study to include neutrino fields, he found that overspinning could still occur, and backreaction was insufficient to prevent this outcome.

Building on this line of inquiry, Yang et al. \cite{yang2025test} revised the WCCC by analyzing KN BHs under perturbations from scalar and neutrino fields. He pointed out that the assumption made by Koray regarding the energy and angular momentum of the test fields absorbed by BH violates the second law of BH thermodynamics.

In this study, we test the WCCC for a general KN-MOG-BH by employing a test particle approach. Section 2 presents the KN-MOG metric along with necessary conditions for violating the WCCC. Section 3 focuses on particular cases — including Reissner–Nordström (RN), Kerr-MOG, extremal RN-MOG and non-extremal RN-MOG black holes —to examine whether the event horizon can vanish for valid ranges of particle parameters. In Section 4, the analysis is extended to the fully general KN-MOG -BH, where a more detailed investigation is carried out for both extremal and non-extremal scenarios regarding potential WCCC violation. The concluding section summarizes the key findings of the work.
\section{Kerr-Newman-MOG spacetime and particle dynamics}
\label{sec:1}
The background geometry of the Kerr-Newman-MOG spacetime is given by  \cite{shaymatov2024kerr}
\begin{equation}\label{metric}
	ds^2 = -\left[dt - a\sin^2\theta\, d\phi\right]^2\frac{\Delta}{\rho^2} + \left[\Delta d\theta^2+dr^2\right]\frac{\rho^2}{\Delta} + \left[\left(a^2 + r^2\right)\,d\phi - a\,dt\right]^2\frac{\sin^2\theta}{\rho^2},
\end{equation}

where 
\begin{equation}\label{delta_rho}
	\Delta = r^2 + a^2 + Q^2 - 2(1+\alpha)r M G_N + (1+\alpha)\alpha M^2 G_N^2, \quad 
	\rho^2 = a^2\cos^2\theta+r^2.
\end{equation}

In the context of MOG, the distinction between the effective gravitational constant $G$ and the standard Newtonian constant $G_N$ is measured using a dimensionless parameter $\alpha$, given by
\begin{equation}
	\alpha = \frac{G}{G_N}-1.
\end{equation}

The angular momentum and Arnowitt-Deser-Misner (ADM) mass of the Kerr-Newman-MOG-BH are given by
\begin{equation}
	J = a \mathcal{M}, \quad \mathcal{M} = M(1+\alpha),
\end{equation}
wher $a$ is the spin parameter and $M$ is the mass of black hole. For simplicity and without affecting generality, we adopt natural units with $G_N = 1$ throughout the analysis.

Under these assumptions, the function $\Delta$ appearing in the metric takes the form
\begin{equation}
	\Delta = a^2+r^2+Q^2- 2r\mathcal{M}+ \mathcal{M}^2\frac{\alpha}{1+\alpha}.  
\end{equation}

The radial positions of the event horizons are given by
\begin{equation}\label{KorayModified}
	r_{\pm} = \mathcal{M} \pm \sqrt{ \frac{ \mathcal{M}^2 - (\alpha + 1)(a^2 + Q^2) }{\alpha+1} }.
\end{equation}

To ensure that an event horizon exists, the discriminant under the square root must be non-negative. This requirement yields the condition
\begin{equation}
	\mathcal{M}^2 \geq (a^2 + Q^2)(\alpha+1),
\end{equation}
where the equality corresponds to extremal black hole, which possesses the maximum possible charge and angular momentum for a given mass.


For the Kerr–Newman–MOG-BH, the first law of BH thermodynamics takes the form \cite{bekenstein1973black, bardeen1973four, needham1980cosmic}
\begin{equation} \label{waq1}
	\delta \mathcal{M} = \Phi \delta Q+ \Omega \delta J+T \delta S,
\end{equation}
where $\delta$ indicates change in the black hole's parameters. The corresponding expressions for the electric potential, angular velocity and temperature are
\begin{equation}\label{waq2}
	\Phi = \frac{Q r_+}{a^2+r_+^2} , \quad \Omega = \frac{a}{a^2+r_+^2}, \quad T =\frac{r_+ - \mathcal{M}}{(a^2+r_+^2)2\pi}.
\end{equation}

Assuming the validity of the weak energy condition, the second law of black hole thermodynamics implies that the entropy $S$ cannot decrease, i.e.
\begin{equation}\label{waq3}
	\delta S \geq 0.
\end{equation}

By substituting equations \eqref{waq1} and \eqref{waq2} into inequality \eqref{waq3} and performing some algebraic simplification, one obtains the following condition
\begin{equation}\label{waq4}
	\delta \mathcal{M}_\mathrm{min} \geq \frac{Q\,\delta Q\,r_++a\,\delta J  }{a^2+r_+^2 }.
\end{equation}

 Therefore, when the black hole is subjected to a small perturbation by a test particle, the resulting changes in its parameters must obey~(\ref{waq4}).

Now, consider a test particle characterized by angular momentum \(L\), charge \(q\) and energy \(E\), falling into KN-MOG black hole. As the particle crosses the outer event horizon, it alters the black hole’s angular momentum, charge and mass. During this process, a portion of the particle’s energy may be radiated away. Therefore, by applying the laws of conservation

\begin{equation}\label{new1}
	\delta \mathcal{M}_\mathrm{min} < E_{\mathrm{min}}, \quad \delta Q = q, \quad \delta J = L
\end{equation}
We consider the scenario where the particle’s initial spin is oriented along the symmetry axis of the black hole. Due to this symmetry, the spin remains aligned with the black hole's axis. As a result, once the particle is absorbed, the black hole maintains its axisymmetric nature. Additionally, any radiation emitted by the particle does not contribute to a loss of angular momentum \cite{wald1974gedanken}. 
By comparing~(\ref{waq4}) and~(\ref{new1}), one obtains a lower bound on the energy of the particle

\begin{equation}\label{new2}
	E_{\mathrm{min}} \geq \frac{aL+q Q r_+}{r_+^2+a^2}.  
\end{equation}
Absorption of a particle by the black hole is possible only if the particle's energy exceeds the minimum threshold specified in~(\ref{new2}). Moreover, prior to the particle's infall into the black hole, the condition for the existence of the event horizon must be satisfied. The condition is expressed as 
\begin{equation}\label{waq5}
	\Delta_\mathrm{1} =  \mathcal{M}_\mathrm{1}^4-\left[\alpha+1\right]J_\mathrm{1}^2-\left[\alpha+1\right]\mathcal{M}_\mathrm{1}^2 Q_\mathrm{1}^2\geq 0,
\end{equation}
 where the equality holds for extremal KN-MOG black hole. Since we are
interested in destroying the event horizon by throwing a particle into it, we must
have
\begin{equation}\label{waq6}
	\Delta_\mathrm{2} = \mathcal{M}_\mathrm{2}^4-\left[\alpha+1\right]J_\mathrm{2}^2-\left[\alpha+1\right]\mathcal{M}_\mathrm{2}^2Q_\mathrm{2}^2<0, 
\end{equation}
where $\mathcal{M}_\mathrm{2} = \mathcal{M} + \delta \mathcal{M}$ represents the final mass (or energy), $J_\mathrm{2} = J + L$ denotes the total angular momentum and $Q_\mathrm{2} = Q + q$ is the total charge of KN-MOG black hole after particle absorption. Upon performing straightforward algebraic manipulations, inequality~(\ref{waq6}) simplifies to

\begin{equation}\label{waq7}
	\scalebox{0.95}{$
		 E_\mathrm{max} < \sqrt{\frac{(1+\alpha)(Q+q)^2}{2} + \sqrt{\left(J + L\right)^2(1+\alpha) + \frac{(1+\alpha)^2(Q+q)^4}{4} - \mathcal{M}}}.
		$}
\end{equation}
Equation~(\ref{waq7}) defines the highest permissible energy for the infalling test particle. When combined with the constraint provided by~(\ref{waq5}), the inequalities~(\ref{new2}) and~(\ref{waq7}) together define the allowed range of the particle’s angular momentum, charge and energy that could potentially lead to a violation of the WCCC. Mathematically

\begin{equation} \label{waq8}
	 E_\mathrm{min}< E <  E_\mathrm{max}.
\end{equation}
Next, we consider a general Kerr–Newman–MOG black hole without relying on any small parameter expansions. In such a case, it becomes more convenient to work with dimensionless quantities. To facilitate this, we introduce the following definitions \cite{khodabakhshi2020cosmic}

\[
\epsilon = \frac{ E}{\mathcal{M}}+1, \quad \eta = \frac{L}{J}+1, \quad \xi = \frac{q}{Q}+1, \quad \gamma = \frac{a^2}{\mathcal{M}^2}, \quad \lambda = \frac{Q^2}{\mathcal{M}^2}.
\]
Here, \( \epsilon, \eta \text{ and } \xi \) represent the dimensionless energy, angular momentum and charge of the infalling particle, respectively. The parameters \( \gamma \) and \( \lambda \) denote the dimensionless rotation and charge parameters of the Kerr–Newman–MOG black hole. Based on these definitions, the conditions for the minimum and maximum energy required for horizon destruction, along with the constraint for the existence of the event horizon, are expressed as follows

\begin{equation}\label{waq11}
	E_\mathrm{min} = (\epsilon - 1) - \frac{(1+\sqrt{\frac{1}{1+\alpha}-\gamma-\lambda})\lambda(\xi-1) + \gamma(\eta - 1)}{2\sqrt{\frac{1}{1+\alpha}-\gamma-\lambda} + \left(\frac{2+\alpha}{1+\alpha} - \lambda\right)} \geqslant 0,
\end{equation}

\begin{equation}\label{waq10}
	E_\mathrm{max} = 2\epsilon^2 - (1+\alpha)\lambda \xi^2 - \sqrt{4(1+\alpha)\gamma\eta^2 + (1+\alpha)^2\lambda^2\xi^4} < 0,
\end{equation}

\begin{equation}\label{waq12}
	1-(\lambda+\gamma)(1+\alpha)\geq 0,
\end{equation}
where equality in equation~(\ref{waq12}) corresponds to the extremal case of the Kerr–Newman–MOG black hole. It is important to note that the charge of the particle relative to the black hole can be either like-signed or oppositely signed, i.e., $\xi > 1$ or $\xi < 1$, respectively. Similarly, the particle may co-rotate or counter-rotate with respect to the black hole, which is reflected by $\eta > 1$ or $\eta < 1$. Furthermore, to validate the particle assumption, we require $|\eta|$, $|\xi|$, and $\epsilon \sim 1$.
 
\section{Constraints on the particle's parameters}

We begin by examining some special cases of black holes.

\subsection{WCCC in RN-MOG and Kerr–MOG-BHs}

In RN-MOG black hole, setting the rotation parameter \(\gamma = 0\) eliminates the need for the angular momentum parameter \(\eta\). Under this simplification, the inequalities \eqref{waq11}–\eqref{waq12} reduce to the following

\begin{equation}\label{waq13}
	E_\mathrm{min} = 4\epsilon^2 + (1+\alpha)^2\lambda^2\xi^4 - 2(1+\alpha)\lambda\xi^2 \geqslant 0,
\end{equation}

\begin{equation}\label{waq14}
	E_\mathrm{max} = (\epsilon - 1) - \frac{\left(1 + \sqrt{\frac{1}{1+\alpha} - \lambda}\right)\lambda(\xi - 1)}{2\sqrt{\frac{1}{1+\alpha} - \lambda} + \left(\frac{2+\alpha}{1+\alpha} - \lambda\right)} < 0,
\end{equation}

\begin{equation}\label{waq15}
	1 - \lambda(1 + \alpha) \geq 0.
\end{equation}

We obtain RN (charged) black hole by setting $\alpha = 0$ in \eqref{waq13}–\eqref{waq15}. Under this assumption, the inequalities \eqref{waq13}–\eqref{waq15} take the following form
\begin{equation}\label{k1}
	(1 + \sqrt{1 - \lambda})(\lambda \xi - 2\epsilon - \lambda + 2) + \lambda(\epsilon - 1) \leq 0,
\end{equation}
\begin{equation}\label{k2}
	\lambda \xi^2 > \epsilon^2,
\end{equation}
\begin{equation}\label{k3}
	1 > \lambda.
\end{equation}
Here, the three-dimensional parameter space is characterized by $\lambda$, which is the charge of the black hole, along with $\epsilon$ and $\xi$, denoting the energy and charge of the particle, respectively. 

\begin{figure}[H]
 	\centering
 	\includegraphics[width=0.40\textwidth]{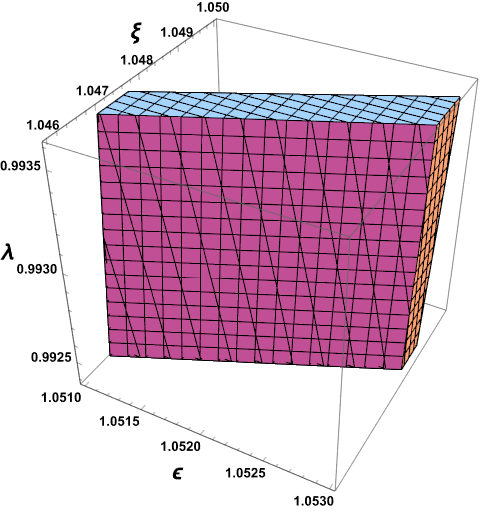} 
 	\caption{Violation of WCCC in three dimensional space of $\epsilon$, $\lambda$ and $\xi$ for non-extremal RN black hole.}
 	\label{fig:ExtremalllRegion0}
 \end{figure}

Plotting the region defined by inequalities \eqref{k1}–\eqref{k2} such that the constraint \eqref{k3} is satisfied, reveals that valid solutions exist within the permissible parameter space (Fig.~\ref{fig:ExtremalllRegion0}).

 It turns out, the WCCC is violated in the case of non extremal RN black hole. It is important to note that Khodabakhshi and Shojai \cite{khodabakhshi2020cosmic} assert that no solutions exist for the corresponding inequalities, our investigation, supported by explicit plotting, reveals the presence of viable solutions within the permissible parameter space. This suggests that their conclusion may not be valid in this context.
 
 For the extremal RN black hole ($\lambda = 1$), the inequalities \eqref{k1}–\eqref{k2} simplify to:
 \begin{equation}
 	\epsilon - \xi \geq 0,
 \end{equation}
 \begin{equation}
 	\xi^2 > \epsilon^2.
 \end{equation}
 It is clear that these two inequalities are inconsistent. As a result, the WCCC remains preserved in the extremal RN black hole case.
 
 Moreover, the WCCC holds for Kerr and extremal Kerr black holes when both $\alpha = 0 = \lambda$ \cite{hubeny1999overcharging,jacobson2009overspinning,huang2022topological}. However, the scenario differs in the case of extremal and near-extremal Kerr-MOG-BH, where violations of the WCCC have been observed \cite{duztacs2020overspinning}.
 
 \subsection{WCCC in extremal RN-MOG-BH}
 
\begin{figure}[H]
	\centering
	
	\begin{subfigure}{0.30\textwidth}
		\centering
		\includegraphics[width=\linewidth]{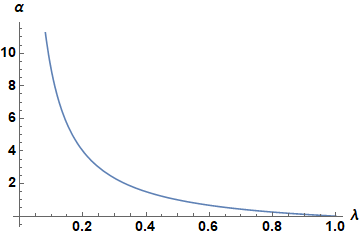}
		\caption{}
		\label{fig:combined_w1}
	\end{subfigure}
	\hfill
	
	\vspace{1em} 
	
	\begin{subfigure}{0.30\textwidth}
		\centering
		\includegraphics[width=\linewidth]{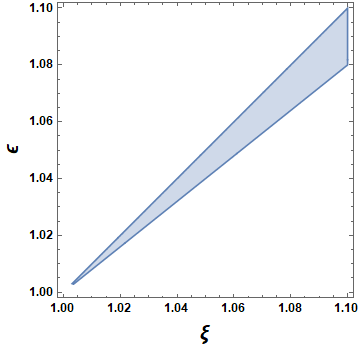}
		\caption{$\alpha = 0.25$, $\lambda = 0.8$}
		\label{fig:combined_w3}
	\end{subfigure}
	\hfill
	\begin{subfigure}{0.30\textwidth}
		\centering
		\includegraphics[width=\linewidth]{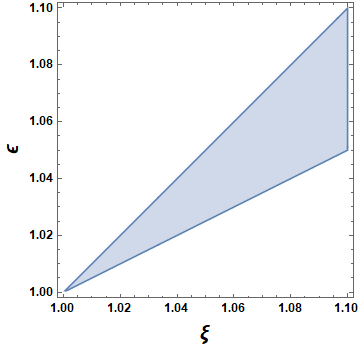}
		\caption{$\alpha = 1$, $\lambda = 0.5$}
		\label{fig:combined_w2}
	\end{subfigure}
	\hfill
	\begin{subfigure}{0.30\textwidth}
		\centering
		\includegraphics[width=\linewidth]{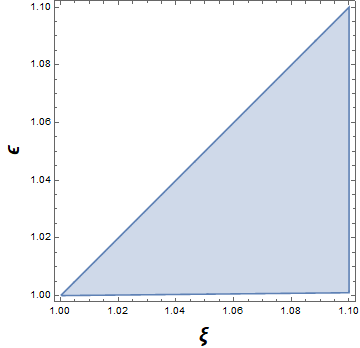}
		\caption{$\alpha = 99$, $\lambda = 0.01$}
		\label{fig:combined_w4}
	\end{subfigure}
	
	\caption{Violation of WCCC in two dimensional space of $\xi$ and $\epsilon$ for specific $\lambda$ and $\alpha$ (obtained from (a)) for the extremal RN-MOG black hole.}
	\label{fig:combined_analysis}
\end{figure}

 Solving inequality \eqref{waq15} for the case $\alpha \neq 0$ yields the condition for the extremal RN-MOG black hole
 \begin{equation}\label{waq16}
 0 < \alpha \leq \frac{1}{\lambda}-1\quad \text{and} \quad 	0 < \lambda < 1  ,
 \end{equation}
 where the equal sign indicates the extremality of RN-MOG black hole. The relation  between $\lambda$ and $\alpha$ is clearly illustrated in Fig.~\ref{fig:combined_w1}.
 
 To investigate the violation of the WCCC, we choose any $\lambda$ from $0 < \lambda < 1$ and assign the corresponding $\alpha = \frac{1}{\lambda}-1$ such that the constraint \eqref{waq15} is fulfilled, we plot inequalities \eqref{waq13} and \eqref{waq14} in the two-dimensional parameter space of $\xi$ and $\epsilon$, identifying the ranges for which the WCCC is violated. The results of this analysis are shown in Fig.~\ref{fig:combined_analysis}
  \protect\subref{fig:combined_w3}-\protect\subref{fig:combined_w4}, which illustrate the WCCC violating ranges of $\xi$ and $\epsilon$ for specified values of $\alpha$ and $\lambda$, indicating the violation of WCCC. One can clearly see from these plots that the admissible range of particle parameters expands as the MOG parameter $\alpha$ increases and the black hole charge $\lambda$ decreases.
   
\subsection{WCCC in non-extremal RN-MOG-BH}

Excluding the equal sign from inequality \eqref{waq16} leaves us with a non-extremal RN-MOG black hole. Choosing any $\lambda$ within the open interval $(0,1)$ and a corresponding value of $\alpha$ satisfying $0 < \alpha < \frac{1 - \lambda}{\lambda}$ such that the condition \eqref{waq15} is fulfilled, we plot inequalities \eqref{waq11} and \eqref{waq10} for $\gamma = 0$ in the three-dimensional parameter space of $\eta$, $\xi$, and $\epsilon$ to determine the regions where the WCCC is violated.

\begin{figure}[H]
	\centering
	\begin{subfigure}{0.32\textwidth}
		\centering
		\includegraphics[width=\textwidth]{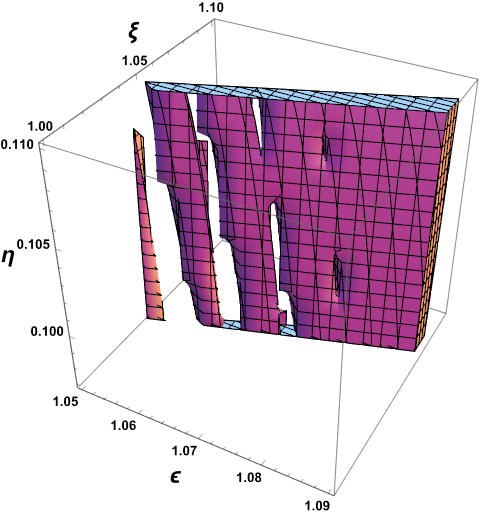}
		\caption{$0 < \alpha < 0.1111, \lambda = 0.9$}
		\label{fig:nonextremal1}
	\end{subfigure}
	\hfill
	\begin{subfigure}{0.32\textwidth}
		\centering
		\includegraphics[width=\textwidth]{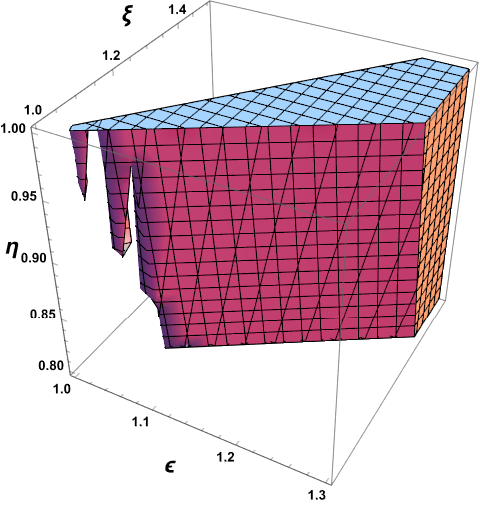}
		\caption{$0 < \alpha < 1, \lambda = 0.5$}
		\label{fig:nonextremal2}
	\end{subfigure}
	\hfill
	\begin{subfigure}{0.32\textwidth}
		\centering
		\includegraphics[width=\textwidth]{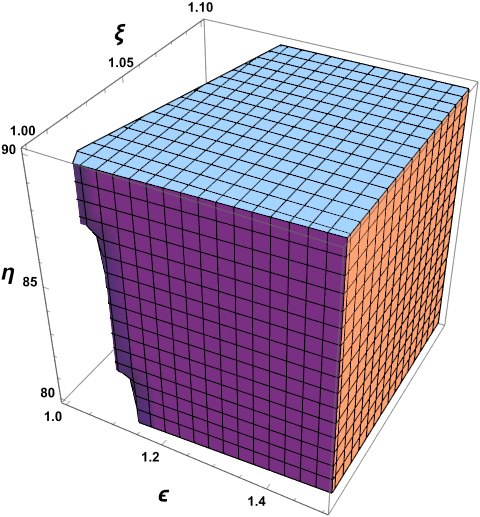}
		\caption{$0 < \alpha < 99, \lambda = 0.01$}
		\label{fig:nonextremal3}
	\end{subfigure}
	
	\caption{Violation of WCCC in three dimensional space of $\xi,\ \eta\ \text{and}\ \epsilon$ for specific $\lambda$ and $\alpha$ for non-extremal RN-MOG black hole.}  
	\label{fig:nonextremal_all}
\end{figure}

The plots in Fig.~\ref{fig:nonextremal_all} \protect\subref{fig:nonextremal1}–\protect\subref{fig:nonextremal3} illustrate the parameter ranges where this violation occurs for selected values of $\lambda$ and $\alpha$. It is evident from these plots that the admissible range of particle parameters expands as the range of the MOG parameter broadens and the black hole’s charge decreases. This analysis confirms that the WCCC is also violated in the non-extremal RN-MOG black hole.

\section{WCCC in Kerr–Newman–MOG-BH}

In this section, we investigate the validity of the WCCC for two specific cases of the KN-MOG black hole.

\subsection{Extremal case}

A KN-MOG-BH becomes extremal when the relation $1 - (\lambda + \gamma)(1 + \alpha) = 0$ holds. To allow a charged and rotating particle to fall into extremal KN-MOG black hole in such a way that it may overspin or overcharge the black hole into a naked singularity, one must ensure that the particle’s parameters are chosen such that the inequalities~\eqref{waq11}–\eqref{waq12} remain satisfied after the encounter.

\begin{figure}[H]
	\centering
	\includegraphics[width=0.32\textwidth]{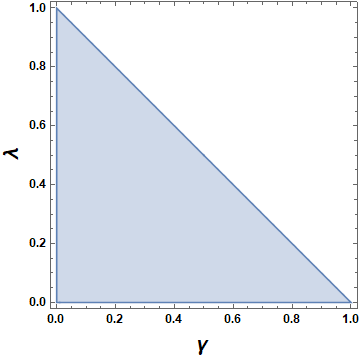} 
	\caption{Two-dimensional parameter space of $\gamma$ and $\lambda$ for $\alpha > 0$}
	\label{fig:ExtremalRegionnn}
\end{figure}

Fig.~\ref{fig:ExtremalRegionnn} gives the region in the two-dimensional parameter space of $\gamma$ and $\lambda$ for which extremality is achieved, given $\alpha = \frac{1}{\gamma + \lambda} - 1 > 0$.

Substituting $\alpha = \frac{1}{\gamma + \lambda} - 1$ into inequalities \eqref{waq11} and \eqref{waq10} for the extremal case, we obtain 

\begin{equation}
	E_\mathrm{min} = \epsilon - 1 + \gamma\left(\epsilon + \lambda(1 - \xi) - \eta\right)\geq 0,
\end{equation}

\begin{equation}
	E_\mathrm{max} = 2\epsilon^2(\gamma + \lambda) - \left[\lambda\xi^2 + \sqrt{\lambda^2\xi^4 +4\gamma\eta^4(\gamma + \lambda)}\right] < 0.
\end{equation}

For extremal KN-MOG black holes, the MOG parameter $\alpha$ must satisfy $\alpha > 0$. This parameter depends on both the spin $\gamma$ and charge parameter $\lambda$. As Fig.~\ref{fig:ExtremalRegionnn} demonstrates, $\alpha$ remains positive throughout the entire domain where $\lambda, \gamma \in (0, 1)$. Selecting any $\lambda$ and $\gamma$ within this region consequently results in $\alpha > 0$.

By selecting values of $\lambda$ and $\gamma$ from the open interval $(0, 1)$ such that the constraint defined by \eqref{waq12} is satisfied, we plot inequalities \eqref{waq11} and \eqref{waq10} for $\alpha > 0$ in the three-dimensional parameter space of $\eta$, $\xi$ and $\epsilon$. This allows us to identify the regions where the WCCC is violated.

\begin{figure}[H]
	\centering
	
	\begin{minipage}{\textwidth}
		\begin{subfigure}{1\textwidth}
			\centering
			\includegraphics[width=\textwidth]{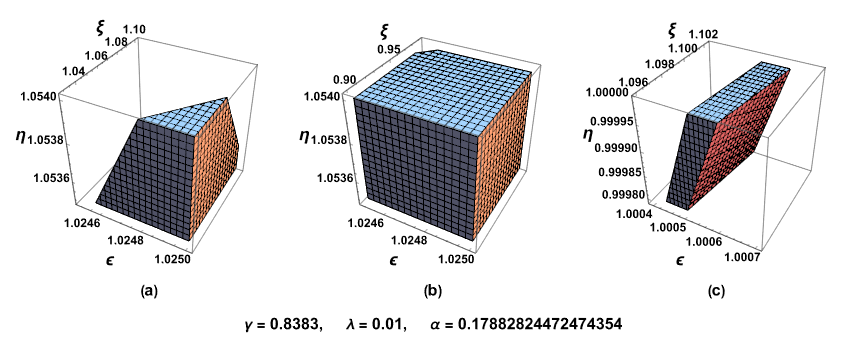}
			\label{fig:knnonextremal1}
		\end{subfigure}
		\hfill
		\begin{subfigure}{1\textwidth}
			\centering
			\includegraphics[width=\textwidth]{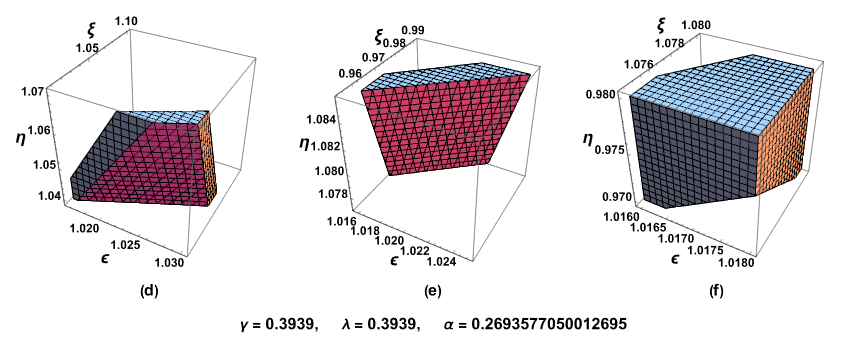}
			\label{fig:knnonextremal2}
		\end{subfigure}
		\hfill
		\begin{subfigure}{1\textwidth}
			\centering
			\includegraphics[width=\textwidth]{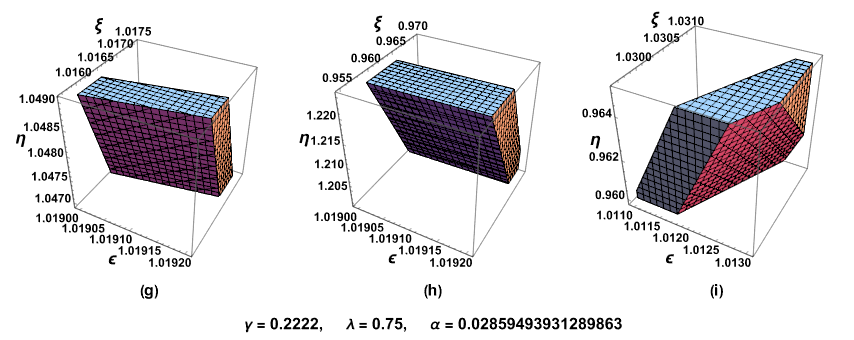}
			\label{fig:knnonextremal3}
		\end{subfigure}
		\caption*{ }
	\end{minipage}
	
	\vspace{1em}
	
	\caption{
		Violation of WCCC in three-dimensional space of $\epsilon,\ \eta\ \text{and}\ \xi$ for specific values of $\alpha,\ \lambda\ \text{and}\ \gamma$ for extremal KN-MOG black hole.
	}
	\label{fig:waq1}
\end{figure}

The results are presented in Fig.~\ref{fig:waq1}, which displays the parameter ranges for which WCCC violation occurs for specific values of $\alpha$, $\lambda$ and $\gamma$ under three scenarios
 \begin{equation}
	\gamma > \lambda, \quad \gamma \sim \lambda, \quad \gamma < \lambda.
\end{equation}
 Since the particle and the black hole may rotate in either the same or opposite directions, the parameters $\xi$ and $\eta$ can slightly exceed or fall below unity.

 Now consider for example, the Fig.~\ref{fig:waq1}(a), the WCCC is not preserved within the following intervals of test particle parameters for $\gamma > \lambda$
 \begin{equation}\label{new3}
 	1.0536 < \eta < 1.0540, \quad 
 	1.0246 < \epsilon < 1.0250, \quad 
 	1.0536 < \xi < 1.0540.
 \end{equation}
 
 When all fundamental constants---Planck's constant $\hbar$, the speed of light $c$, and the gravitational constant $G$---are reinstated, the inequalities in \eqref{new3} determine the lower and upper bounds for the energy and dimensionless spin of the test particle. These bounds are summarized in Table~\ref{tab:Extremal-KN-MOG-BHH}, which contrasts the energy and angular momentum parameters in the extremal cases of both the KN black hole and the KN-MOG black hole.
 
 Building upon this analysis, Table~\ref{tab:Extremal-KN-MOG-BH-GRSS} extends the model to the Milky Way’s well-known stellar-mass black hole GRS~1915+105, whose mass lies within the range $\mathcal{M} \sim 10$--$18\mathcal{M}_{\odot}$ and whose dimensionless spin parameter satisfies $\frac{c J}{G\mathcal{M}^2_{\odot}} \sim 0.8$--$1$. It is evident that the inclusion of the MOG parameter $\alpha$ broadens the admissible energy and angular momentum range of the particle, enabling the violation of the WCCC for GRS~1915+105 in the extremal KN-MOG black hole scenario.
 
\begin{table}[h]
	\centering
	\renewcommand{\arraystretch}{2} 
	\begin{tabular}{|p{3cm}<{\centering}|p{8cm}|} 
		\hline
		& \textbf{Energy/Angular momentum of particle} \\  
		\hline
		\multirow{2}{*}{KN black hole \citep{khodabakhshi2020cosmic}} & $0.00128\frac{\mathcal{M}}{\mathcal{M}_{\odot}}<\frac{E}{c^2\mathcal{M}_{\odot}}<0.0013\frac{\mathcal{M}}{\mathcal{M}_{\odot}}$ \\  
		\hhline{~-}
		& $0.0008698\left(\frac{\mathcal{M}}{\mathcal{M}_{\odot}}\right)^2<\frac{L }{\mathcal{M}^2_{\odot}G/c}<0.0008881\left(\frac{\mathcal{M}}{\mathcal{M}_{\odot}}\right)^2$  \\  
		\hline
		\multirow{2}{*}{KN-MOG black hole} & $0.0246\frac{\mathcal{M}}{\mathcal{M}_{\odot}}<\frac{E}{c^2\mathcal{M}_{\odot}}<0.025\frac{\mathcal{M}}{\mathcal{M}_{\odot}}$  \\  
		\hhline{~-}
		& $0.0536\left(\frac{\mathcal{M}}{\mathcal{M}_{\odot}}\right)^2<\frac{L}{\mathcal{M}^2_{\odot}G/c}<0.054\left(\frac{\mathcal{M}}{\mathcal{M}_{\odot}}\right)^2$\\  
		\hline
	\end{tabular}
	\caption{The lower and upper bounds of energy and angular momentum of the particle to destroy the horizon of KN and KN-MOG black holes in extremal case.}
	\label{tab:Extremal-KN-MOG-BHH}
\end{table}
\begin{table}[h]
	\centering
	\renewcommand{\arraystretch}{2} 
	\begin{tabular}{|p{3cm}<{\centering}|p{8cm}|} 
		\hline
		& \textbf{Energy/Angular momentum of particle} \\  
		\hline
		\multirow{2}{*}{KN black hole \citep{khodabakhshi2020cosmic}} & $0.0128<\frac{E}{c^2\mathcal{M}_{\odot}}<0.0234$ \\  
		\hhline{~-}
		& $0.08698<\frac{L}{\mathcal{M}^2_{\odot}G/c}<0.2877$ \\  
		\hline
		\multirow{2}{*}{KN-MOG black hole} & $0.246<\frac{E}{c^2\mathcal{M}_{\odot}}<8.1$ \\  
		\hhline{~-}
		& $5.36<\frac{L}{\mathcal{M}^2_{\odot}G/c}<17.496$\\  
		\hline
	\end{tabular}
	\caption{The lower and upper bounds of energy and angular momentum of the particle to destroy the horizon of stellar-mass black hole GRS 1915+105 in extremal case. }
	
	\label{tab:Extremal-KN-MOG-BH-GRSS}
\end{table}


\subsection{Non-extremal case}

 The KN-MOG black hole is non-extremal if 
 \( 1 - (\lambda + \gamma)(1 + \alpha) > 0 \). From this constraint, we obtain
\begin{equation}\label{KN-MOGg}
	0 < \lambda < 1, \quad 0 < \gamma < - \lambda+1, \quad 0 < \alpha < -1+(\lambda+\gamma)^{-1}.
	.
\end{equation}
By selecting $\lambda$ from the range $0 < \lambda < 1$, we obtain open intervals for each $\gamma$ and $\alpha$. This behavior differs from the extremal case, where $\alpha$ is uniquely determined for specified $\gamma$ and $\lambda$. We consider three special cases in the light of \eqref{KN-MOGg}
\begin{equation}
	\left.
	\begin{aligned}
		\lambda &= 0.01,   &\quad 0 < \gamma < 0.99,   &\quad 0 < \alpha < 0.178828 \\
		\lambda &= 0.3939, &\quad 0 < \gamma < 0.6061, &\quad 0 < \alpha < 0.269358 \\
		\lambda &= 0.2222, &\quad 0 < \gamma < 0.7778, &\quad 0 < \alpha < 1.68673
	\end{aligned}
	\right\}
\end{equation}

By choosing values of $\gamma$ and $\alpha$ corresponding to each $\lambda$ from the above cases such that the condition $1 - (1 + \alpha)(\gamma + \lambda) > 0$ is satisfied, we plot inequalities \eqref{waq11} and \eqref{waq10} in the three-dimensional space of parameters $\epsilon$, $\xi$ and $\eta$ to identify the regions in which the WCCC is violated.

\begin{figure}[H]
	\centering
	
	\begin{minipage}{\textwidth}
		\begin{subfigure}{1\textwidth}
			\centering
			\includegraphics[width=\textwidth]{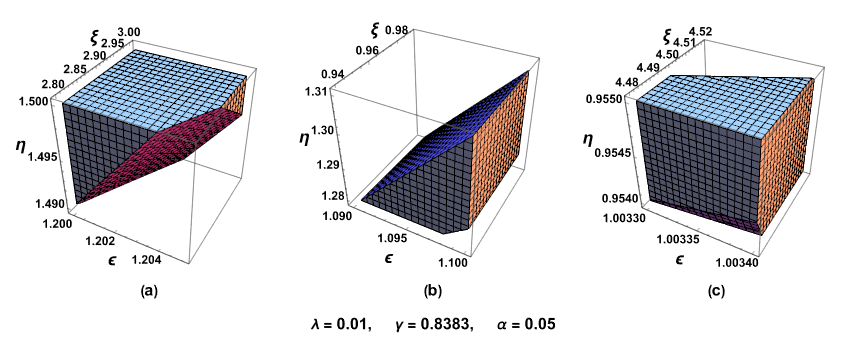}
			\label{fig:knnonextremal1}
		\end{subfigure}
		\hfill
		\begin{subfigure}{1\textwidth}
			\centering
			\includegraphics[width=\textwidth]{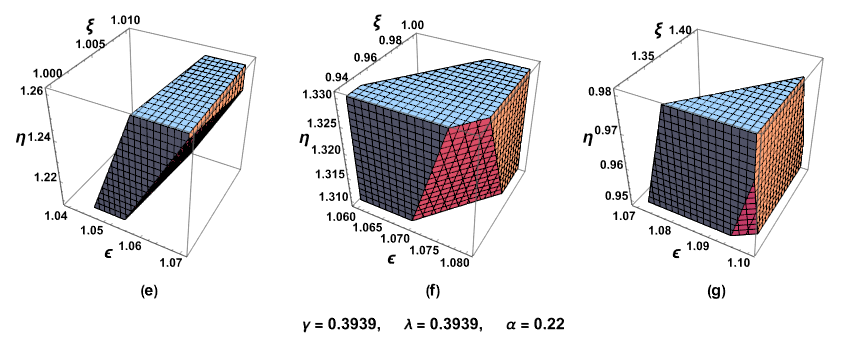}
			\label{fig:knnonextremal2}
		\end{subfigure}
		\hfill
		\begin{subfigure}{1\textwidth}
			\centering
			\includegraphics[width=\textwidth]{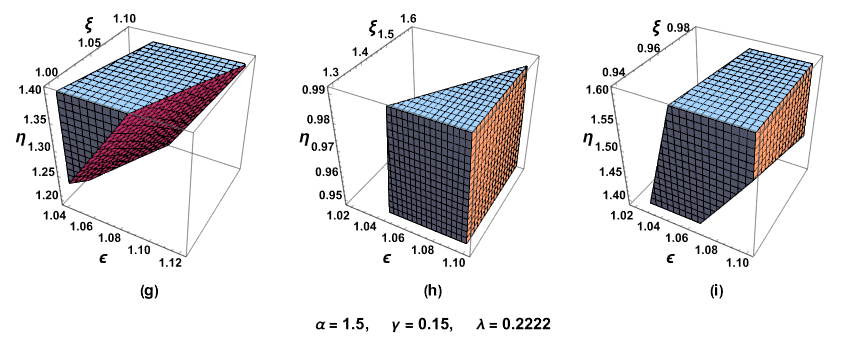}
			\label{fig:knnonextremal3}
		\end{subfigure}
		\caption*{ }
	\end{minipage}
	
	\vspace{1em}
	
	\caption{
		Violation of WCCC in three-dimensional space of $\epsilon,\ \eta\ \text{and}\ \xi$ for specific values of $\alpha,\ \lambda\ \text{and}\ \gamma$ for non-extremal KN-MOG black hole.
	}
	\label{fig:knnonextremal_all}
\end{figure}

Fig.~\ref{fig:knnonextremal_all} illustrates the violation regions for selected values of $\alpha$, $\lambda$, and $\gamma$ under three different comparative settings\begin{equation}
	\gamma > \lambda, \quad \gamma \sim \lambda, \quad \gamma < \lambda.
\end{equation}

As the test particle and black hole may rotate in the same or opposite directions, allowing $\xi$ and $\eta$ to slightly deviate from unity. For example, in Fig.~\ref{fig:knnonextremal_all}(a), taking $\alpha = 0.05$, $\lambda = 0.01$, and $\gamma = 0.8383$, the WCCC is violated within the following parameter ranges
\begin{equation}\label{new4}
	1.490 < \eta < 1.500, \quad 1.200 < \epsilon < 1.205, \quad 2.80 < \xi < 3.0
\end{equation}

The bounds obtained from \eqref{new4} determine the allowed range of energy and dimensionless spin for the particle. These ranges are detailed in Table~\ref{tab:Non-Extremal-KNN-MOG-BH}, where the results are compared for non-extremal KN and KN-MOG black holes.

Table~\ref{tab:Non-Extremal-KN-MOGG-BH-GRS} further showcases the implementation of the model on the stellar-mass black hole GRS~1915+105, which lies within the Milky Way.
 The inclusion of the MOG parameter $\alpha$ leads to an extended range of energy and angular momentum, allowing GRS~1915+105 to potentially violate the WCCC in the non-extremal scenario.


\begin{table}[h]
	\centering
	\renewcommand{\arraystretch}{2} 
	\begin{tabular}{|p{3cm}<{\centering}|p{8cm}|} 
		\hline
		& \textbf{Energy/Angular momentum of particle} \\  
		\hline
		\multirow{2}{*}{KN black hole \citep{khodabakhshi2020cosmic}} & $0.039\frac{\mathcal{M}}{\mathcal{M}_{\odot}}<\frac{E}{c^2\mathcal{M}_{\odot}}<0.042\frac{\mathcal{M}}{\mathcal{M}_{\odot}}$ \\  
		\hhline{~-}
		& $0.078\left(\frac{\mathcal{M}}{\mathcal{M}_{\odot}}\right)^2<\frac{L}{\mathcal{M}^2_{\odot}G/c}<0.084\left(\frac{\mathcal{M}}{\mathcal{M}_{\odot}}\right)^2$ \\  
		\hline
		\multirow{2}{*}{KN-MOG black hole} & $0.2\frac{\mathcal{M}}{\mathcal{M}_{\odot}}<\frac{E}{c^2\mathcal{M}_{\odot}}<0.205\frac{\mathcal{M}}{\mathcal{M}_{\odot}}$ \\  
		\hhline{~-}
		& $4.9\times10^{-1}\left(\frac{\mathcal{M}}{\mathcal{M}_{\odot}}\right)^2<\frac{L}{\mathcal{M}^2_{\odot}G/c}<5\times10^{-1}\left(\frac{\mathcal{M}}{\mathcal{M}_{\odot}}\right)^2$\\  
		\hline
	\end{tabular}
	\caption{The lower and upper bounds of energy and angular momentum of the particle to destroy the horizon of KN and KN-MOG black holes in non-extremal case.}
	\label{tab:Non-Extremal-KNN-MOG-BH}
\end{table}
\begin{table}[h]
	\centering
	\renewcommand{\arraystretch}{2} 
	\begin{tabular}{|p{3cm}<{\centering}|p{8cm}|} 
		\hline
		& \textbf{Energy/Angular momentum of particle} \\  
		\hline
		\multirow{2}{*}{KN black hole \citep{khodabakhshi2020cosmic}} & $0.39<\frac{E}{c^2\mathcal{M}_{\odot}}<0.756$ \\  
		\hhline{~-}
		& $7.8<\frac{L}{\mathcal{M}^2_{\odot}G/c}<27.216$ \\  
		\hline
		\multirow{2}{*}{KN-MOG black hole} & $2<\frac{E}{c^2\mathcal{M}_{\odot}}<3.69$ \\  
		\hhline{~-}
		& $49<\frac{L}{\mathcal{M}^2_{\odot}G/c}<162$\\  
		\hline
	\end{tabular}
	\caption{The lower and upper bounds of energy and angular momentum of the particle to destroy the horizon of stellar-mass black hole GRS 1915+105 in non-extremal case.}
	
	\label{tab:Non-Extremal-KN-MOGG-BH-GRS}
\end{table}

\vspace{6pt}

\section{Conclusion}
The possible breakdown of the WCCC was examined through an analysis of test particle dynamics in the KN-MOG-BH, explicitly neglecting backreaction effects. Our comprehensive analysis elucidates that the WCCC can indeed be violated within this framework, a finding which robustly extends to encompass both extremal and non-extremal cases of the KN-MOG black hole. Remarkably, the parameter space allowing for such violations is significantly wider than that reported in the standard KN black hole with $\alpha = 0$ \cite{khodabakhshi2020cosmic}, highlighting the profound impact of the MOG parameter on the black hole’s horizon stability.

We have systematically identified the admissible ranges of the particle’s energy, charge and angular momentum that result in the violation of the WCCC. These findings are depicted in Figs.~\ref{fig:waq1} and \ref{fig:knnonextremal_all}. The occurrence of WCCC violation is shown to depend critically on three factors: the MOG parameter $\alpha$, the charge $\lambda$ and the angular momentum $\gamma$ of the black hole, whether the black hole and the infalling particle rotate in the opposite or same directions $(\eta < 1 \ \text{or} \ \eta > 1)$ and whether their charges have opposite or same signs $(\xi < 1 \ \text{or} \ \xi > 1)$.

Figs.~\ref{fig:waq1} and \ref{fig:knnonextremal_all} also reveal a notable pattern, WCCC violation is more likely when the test particle has small energy, charge or angular momentum. More importantly, increasing values of particle parameters does not lead to a continuous range that permits WCCC violation; instead, the conditions become more restrictive, requiring greater precision in parameter selection.

Lastly, we extended our analysis to GRS~1915+105, a stellar mass BH in the Milky Way by evaluating the lower and upper bounds of the dimensionless particle’s energy and angular momentum required to induce a violation of the WCCC. The corresponding results for the extremal case are summarized in Tables~\ref{tab:Extremal-KN-MOG-BHH}–\ref{tab:Extremal-KN-MOG-BH-GRSS}, while those for the non-extremal case are provided in Tables~\ref{tab:Non-Extremal-KNN-MOG-BH}–\ref{tab:Non-Extremal-KN-MOGG-BH-GRS}.

\bibliographystyle{plainnat}
\bibliography{sn-bibliography}

\end{document}